\documentclass[12pt]{iopart}
\usepackage{graphicx}
\usepackage{orcidlink}
\usepackage{hyperref}

\begin{document}

\title[Pinning landscape in CaKFe$_4$As$_4$ under pressure]{Competition between the modification of intrinsic superconducting properties and the pinning landscape under external pressure in CaKFe$_4$As$_4$ single crystals.}

\author{  Shuyuan Huyan$^1$\,\orcidlink{0000-0003-0999-2440},  Nestor Haberkorn$^2$\,\orcidlink{0000-0002-5261-1642},  Mingyu Xu$^1$$^{\#}$\,\footnote[0]{$^{\#}$Current address: Department of Chemistry Michigan State University, East Lansing, MI 48824, USA}\orcidlink{0000-0001-7138-6009},Paul C Canfield$^1$\orcidlink{0000-0002-7715-0643} and Sergey L  Bud'ko$^1$$^*$\footnote[0]{$^*$Author to whom any correspondence should be addressed.}\orcidlink{0000-0002-3603-5585}}

\address{$^1$Ames National Laboratory and Department of Physics and Astronomy, Iowa State University, Ames, Iowa 50011, USA}
\address{$^2$Centro At\'omico Bar\'iloche and Instituto Balseiro, CNEA and Consejo de Investigaciones Cient\'ficas y T\'ecnicas Av. E. Bustillo 9500, R8402AGP, S. C. Bar\'iloche, RN,
Argentina}
\ead{budko@ameslab.gov}
\vspace{10pt}

\begin{abstract}
Measurements of low field magnetization, trapped flux magnetization and 5~K flux creep in single crystal of CaKFe$_4$As$_4$ under pressure up to 7.5~GPa in a diamond pressure cell are presented. The observed evolution of the temperature dependence of the self-field critical current and slowing down of the base temperature flux creep rate are explained within the two sources of pinning hypothesis involving presence of CaFe$_2$As$_2$ intergrowths suggested in the literature. Above the half collapsed tetragonal structural transition under pressure, where superconductivity is non-bulk or absent, critically diminished or no diamagnetism and flux trapped magnetization were observed.

\end{abstract}

%
%
\submitto{\SUST}
%
%
%

\section{Introduction}

Among Fe - based superconductors CaKFe$_4$As$_4$  (a member of broader $Ae$$A$Fe$_4$As$_4$ family \cite{iyo16a} where $Ae$ is alkaline earth and $A$ - alkaline metal) is rare example of a {\it stoichiometric} superconductor with a $T_c \sim 35$~K, upper critical field $H_{c2} \sim 900 - 1000$~kOe and no other, structural or magnetic, transitions observed. \cite{mei16a} Transition metal substitution in CaKFe$_4$As$_4$ causes the emergence of hedgehog-spin-vortex-crystal type antiferromagnetic order that competes with superconductivity. \cite{mei18a,xum22a,xum23a} From an applied perspective, due to its relatively high $T_c$ among Fe - based superconductors, low upper critical field anisotropy $\gamma$, and the aforementioned large $H_{c2}$, CaKFe$_4$As$_4$ appears as a potential candidate for use in superconducting wires.  \cite{sin18a,pyo19a,ish19a,hab19a,hab20a}  A notable feature of CaKFe$_4$As$_4$ single crystals is their unusual increase in vortex pinning, which manifests as a peak in the critical current density ($J_c$) with rising temperature.  \cite{sin18a,pyo19a,ish19a,hab19a} This behavior is attributed to strained CaFe$_2$As$_2$ - rich stacking faults that are superconducting at low temperatures but transition to a normal state between 10 and 20 K, resulting in increased vortex pinning. \cite{ish19a}  Consequently, this system allows for observing how small variations in vortex pinning energies, which compete with thermal fluctuations causing depinning, contribute to anomalous changes in $J_c$ with temperature increases and magnetic field variations. 

Whereas introducing defects through irradiation typically modifies the pinning landscape, \cite{hab20a} another way to tune vortex pinning in CaKFe$_4$As$_4$ is by using external parameters such as pressure. \cite{bud93a,bud93b,jun15a} Given the superconducting nature of nanometric crystalline defects, external pressure is expected to influence both the bulk superconducting properties and the characteristics of these pinning centers. The overall effect of pressure \cite{kal17a} on the properties of CaKFe$_4$As$_4$ is rather unusual: initially $T_c$ decreases weakly with pressure.  Above 4~GPa a first order structural transition to a half-collapsed tetragonal (hcT) phase occurs, characterized by As-As bonding forming across the Ca layer, leading to an abrupt loss of {\it bulk} superconductivity. With further pressure increase, at 11 - 12~GPa, the second hcT transition develops, where  As-As bonding forms across the K layer, \cite{kal17a,bor18a,wan23a} and superconductivity is not detected as well. 

For superconducting defects with $T_c$ between 10 and 20 K, which are likely associated with CaFe$_2$As$_2$ and may differ significantly from bulk properties due to stress and proximity effects, \cite{ish19a} moderate non-hydrostatic pressure is known to stabilize superconductivity and suppress structural transitions. This results in a dome - like effect where $T_c$ is initially suppressed as pressure exceeds 0.6~GPa, a range much smaller compared to CaKFe$_4$As$_4$. \cite{tor08a}

Traditionally, pressure studies of $J_c$ are conducted using either current - voltage $(I-V)$ curves (see e.g. Refs. \cite{bud93a,bud93b,jun15a}), or via magnetization loops along with the Bean's critical state model. \cite{bea62a,bea64a} The latter approach for Fe - based superconductors is briefly reviewed in Ref. \cite{san21a}. Measurements of magnetization loops on the sample placed in a large-volume pressure cell pose significant challenges, primarily due to the need to subtract the background signal associated with the pressure cell. Whereas this can be managed reasonably well with well - designed piston - cylinder cell (like the HMD cell \cite{hmd} or similar) with an ample space for a sample and maximum pressure 1.2 - 1.5~GPa, the accurate and reliable magnetization loops measurements are extremely difficult in a diamond anvil cell (DAC). \cite{dro15a,min22a,ere22a}.  This complication is less critical when focusing solely on self-field critical current density.  It was demonstrated \cite{min23a}  that, unlike traditional dc magnetization measurements, the  trapped flux magnetization data are almost unaffected by the background signal of the diamond anvil cell due to minimal external magnetic fields involved in this measurement protocol. Thus, the same Bean's model (with an additional factor of two) can be used to obtain the critical current density (assuming moderate or strong pinning and reasonably slow relaxation). At ambient pressure \cite{bud23a} the values of self-flux $J_c(T)$ obtained from trapped flux magnetization and magnetization loops for CaKFe$_4$As$_4$ (as well as CaK(Fe$_{1-x}$Mn$_x$)$_4$As$_4$ and MgB$_2$) crystals  are very close. Using the same arguments, the self-field creep measurements should not be measurably affected by the signal from the pressure cell, provided the cell does not have ferro-,  ferrimagnetic or spin-glass-like magnetic signature.

In this work we aim to extend the pressure work on CaKFe$_4$As$_4$ to address the evolution of critical currents and vortex dynamics, focusing on their impact on both the bulk superconducting properties and the pinning landscape. Building on the results from Ref. \cite{bud24a} we study self-field $J_c(T)$ (via trapped flux magnetization measurements) and flux creep in CaKFe$_4$As$_4$ single crystal as a function of pressure up to $\sim 7.5$~GPa in a diamond anvil cell. This pressure range encompasses both the bulk superconductivity of CaKFe$_4$As$_4$ and non-bulk (filamentary) superconductivity in the hcT phase, thus probing the trapped flux magnetization response of a filamentary superconductor sample. Additionally, we discuss how pressure affects superconducting defects with moderate $T_c$, particularly focusing on changes in self-field $J_c(T)$ and modifications in flux creep at low temperatures.

\section{Experimental details}

The experimental methods in this work are very similar to those detailed in Ref. \cite{bud24a}, yet to spare the reader's effort to look for reference, we repeat relevant part of them here.

Single crystals of CaKFe$_4$As$_4$  with sharp superconducting transitions ($T_c = 34.7$~K) were grown using high-temperature solution growth. \cite{mei16a,mei17a} These crystals grow as mirrorlike, metallic, micaceous plates with the crystallographic $c$ axis perpendicular to the plate surface (as determined by X-ray diffraction). The sample used in the measurements was a thin plate with the $c$  axis perpendicular to the plate and approximately of a cuboid shape with some irregularities. The width and the length of the sample were measured using an optical microscope as $\sim 0.25 \times 0.26$~mm$^2$ respectively. Note that the sample was notably larger than in Ref. \cite{bud24a}. Still, neither the thickness nor the mass were determined due to resolution of available instruments and a concern for keeping the sample intact. Using the same approach as before \cite{bud24a} we estimated the thickness as $\sim 0.1$~mm and the mass as $\sim 35~ \mu$g. The CaKFe$_4$As$_4$ crystal used in this work was cleaved and cut from one of the larger crystals, that showed no detectable presence of secondary phases in magnetization measurements (see Ref. \cite{mei17a}  for the details of how the samples are pre-selected). The sample in the DAC was oriented with $H \| c$.

The DC magnetization measurements under high pressure were performed in a Quantum Design MPMS-classic SQUID magnetometer, with the standard for this unit scan length of 60~mm.  To ensure better thermalization of the DAC, for each temperature point, after instrument temperature stabilization a delay of 45~sec. was implemented. A 300~sec, wait time was used after the change of magnetic field. The same wait / delay times were used  in our previous measurements with the same DAC in the same MPMS-classic  \cite{bud24a,huy24a}.  For more convenient temperature control, in the trapped flux measurements with DAC, we used $T = 5$~K as a base temperature. A commercial DAC (Almax - easyLab Mcell Ultra \cite{mcell}) with a pair of 700-$\mu$m-diameter culet-sized diamond anvils and tungsten gasket with 400-$\mu$m-diameter hole, was used. Nujol mineral oil, that solidifies at $\sim 1.3$~GPa at room temperature, served as the pressure medium.  \cite{ada82a} The pressure at room temperature was measured by the $R_1$ fluorescence line of a ruby ball. \cite{she20a} The background signal of the DAC with gasket, ruby ball, and pressure medium but without sample was measured under 2~GPa using the same protocols as in the measurements with the sample. 

The magnetization of the sample was analyzed using a  point-by-point subtraction of the SQUID response with/without the sample, and then a dipole fitting on the resulting curve, following Ref. \cite{coa20a}. 

$M(T)$ measurements using zero-field-cooled (ZFC) protocol were performed at 200~Oe applied field. Temperature-dependent trapped flux magnetization as well as magnetization relaxation at 5~K were measured after cooling the sample from above the $T_c$ to 5~K in an applied field of $H_M = 20$~kOe and decreasing the field to a nominal zero value. Note that with $H_M = 20$~kOe  in field-cooled protocol, it ensured that the CaKFe$_4$As$_4$ crystal is in the critical state with vortices fully penetrated inside the sample. \cite{bud23a}.

\section{Results}

Fig. \ref{F1} shows ZFC $M(T)$ data measured at different pressures. For $0 \leq P \leq 4.6$~GPa the 5~K (base temperature) diamagnetic signal associated with the superconducting state is effectively the same, as well as the width of the transition, except for $P = 4.6$~GPa the transition is somewhat broadened. At 5.3~GPa the transition is wide with a possibility of two steps, at 6.1~GPa the diamagnetic signal at 5~K is only 5-6\% of that at  $0 \leq P \leq 4.6$~GPa and finally at 7.5 GPa no anomaly is discerned. As discussed below, the dependence of $T_c$ on pressure is similar to that previously reported in Ref. \cite{kal17a}, where the dramatic decrease in the superconducting volume is related to the hcT transition.

\begin{figure}
\begin{center}
\includegraphics[angle=0,width=150mm]{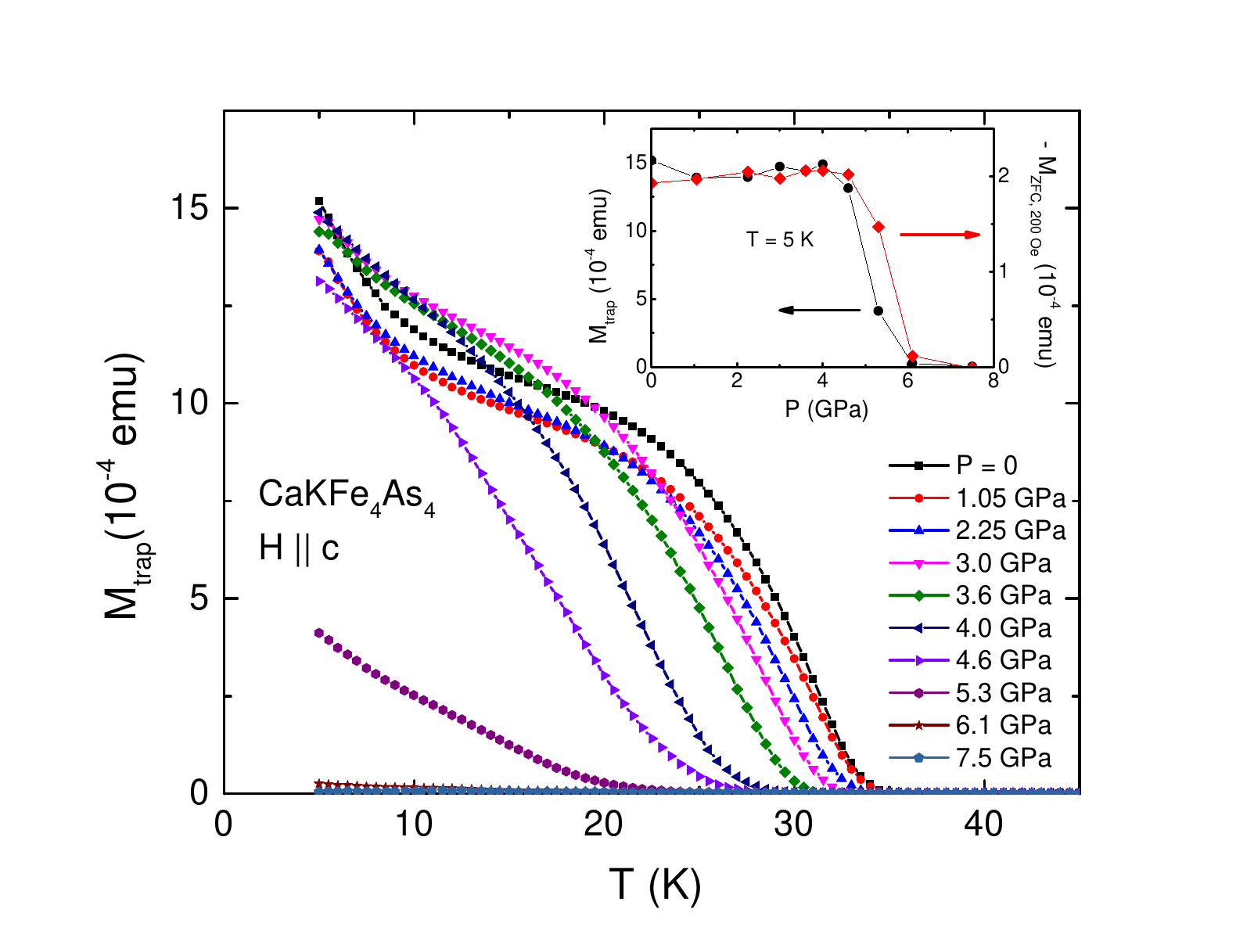}
\end{center}
\caption{(color online) Temperature dependent trapped flux magnetization, $M_{trap}(T)$ of  CaKFe$_4$As$_4$ crystal under pressure measured after coooling the sample in  20~kOe magnetic field applied along the $c$-axis and reducing field to zero. Inset: left axis, black circles $M_{trap}(5~\textrm{K})$ as a function of pressure; right axis, red rhombi $- M(5~\textrm{K})$ measured at 200 Oe with ZFC protocol.} \label{F2} 
\end{figure}

Trapped flux magnetization (after cooling from above $T_c$ to 5~K in 20~kOe magnetic field and reducing field to zero) is shown in Fig. \ref{F2}. Two interesting features can be observed in these data. First, although $T_c$ decreases with pressure, $M_{trap}(5~\textrm{K})$ (see inset to Fig. \ref{F2}) does not change substantially up to 4.6~GPa. $M_{trap}(T)$ then becomes significantly smaller at 5.3~GPa and is basically zero for 6.1 and 7.5 GPa. This behavior is qualitatively similar to the pressure dependence of the 5~K magnetization; both are plotted in the inset of Fig. \ref{F2}. Second, the presence of a broad hump, or shoulder, at $20 - 25$~K in $M_{trap}(T)$ for $0 \leq P \leq 2.25$~GPa is suppressed for $3 \leq P \leq 4.6$~GPa, and it is not observed at extreme pressures where the superconductivity is most likely only filamentary. This change in the shape of $M(T)$ ($\propto J_c(T)$) under pressure suggests changes in the pinning landscape.

\begin{figure}
\begin{center}
\includegraphics[angle=0,width=110mm]{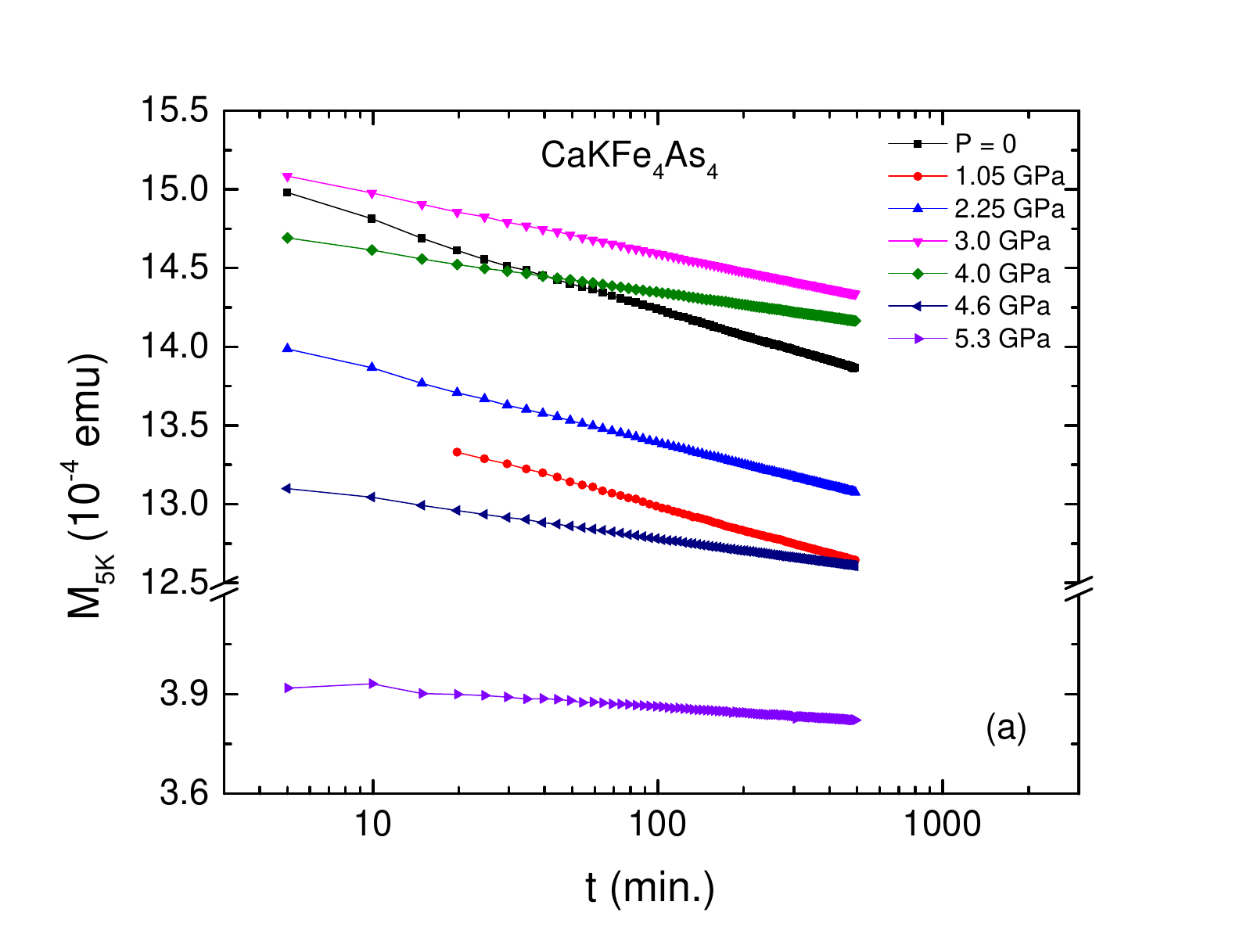}
\includegraphics[angle=0,width=110mm]{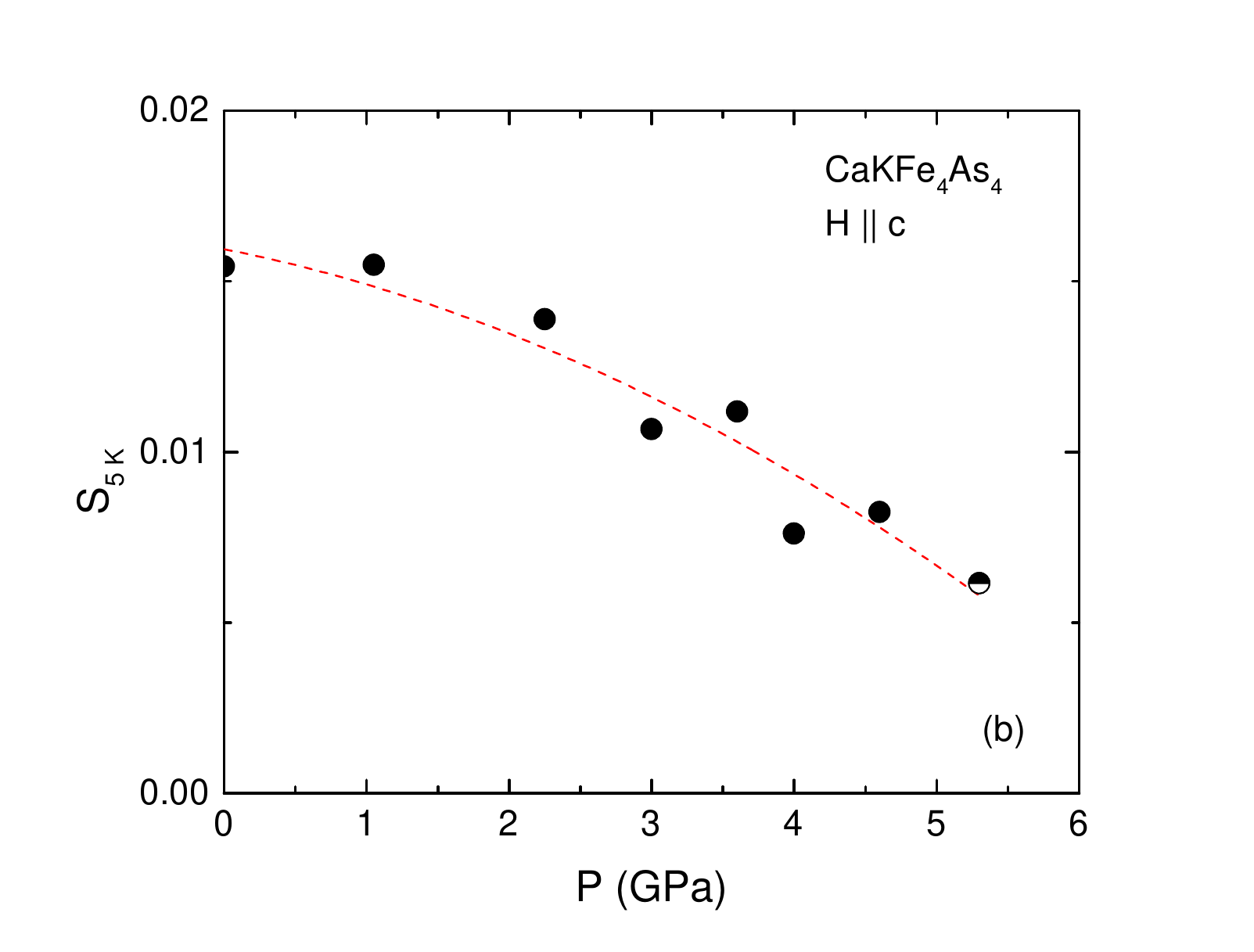}
\end{center}
\caption{(color online) (a) Semi-{\it log} plot of the time dependent trapped flux magnetization of CaKFe$_4$As$_4$ crystal at $T = 5$~K under pressure up to 5.3~GPa measured after coooling the sample in  20~kOe magnetic field applied along the $c$-axis and reducing field to zero. (b) Creep rate at 5~K at different pressures ($S = - 1/M(0) \times dM/d ln(t)$). Dashed line is a guide to the eye. Half-filled symbol corresponds to the data showing potentially not bulk transition. } \label{F3} 
\end{figure}

To analyze the impact of the possible change in the vortex pinning landscape and flux pinning mentioned above, we additionally performed relaxation measurements. Time-dependent magnetization data at 5~K at different pressures after cooling in 20 kOe and reducing the magnetic field to zero are presented in Fig. \ref{F3}(a). As expected,  \cite{yes96a} the data are linear on a semi-{\it log} plot.  As a general trend, it appears that the slope of these lines decreases with increasing pressure. This observation is confirmed on Fig. \ref{F3}(b), where creep rate, $S = - 1/M(0) \times dM/d ln(t)$  ($t$ is time, $M(0)$ is magnetic moment at $t = 0$) at 5~K is plotted as a function of pressure.

\section{Discussion}

\begin{figure}
\begin{center}
\includegraphics[angle=0,width=150mm]{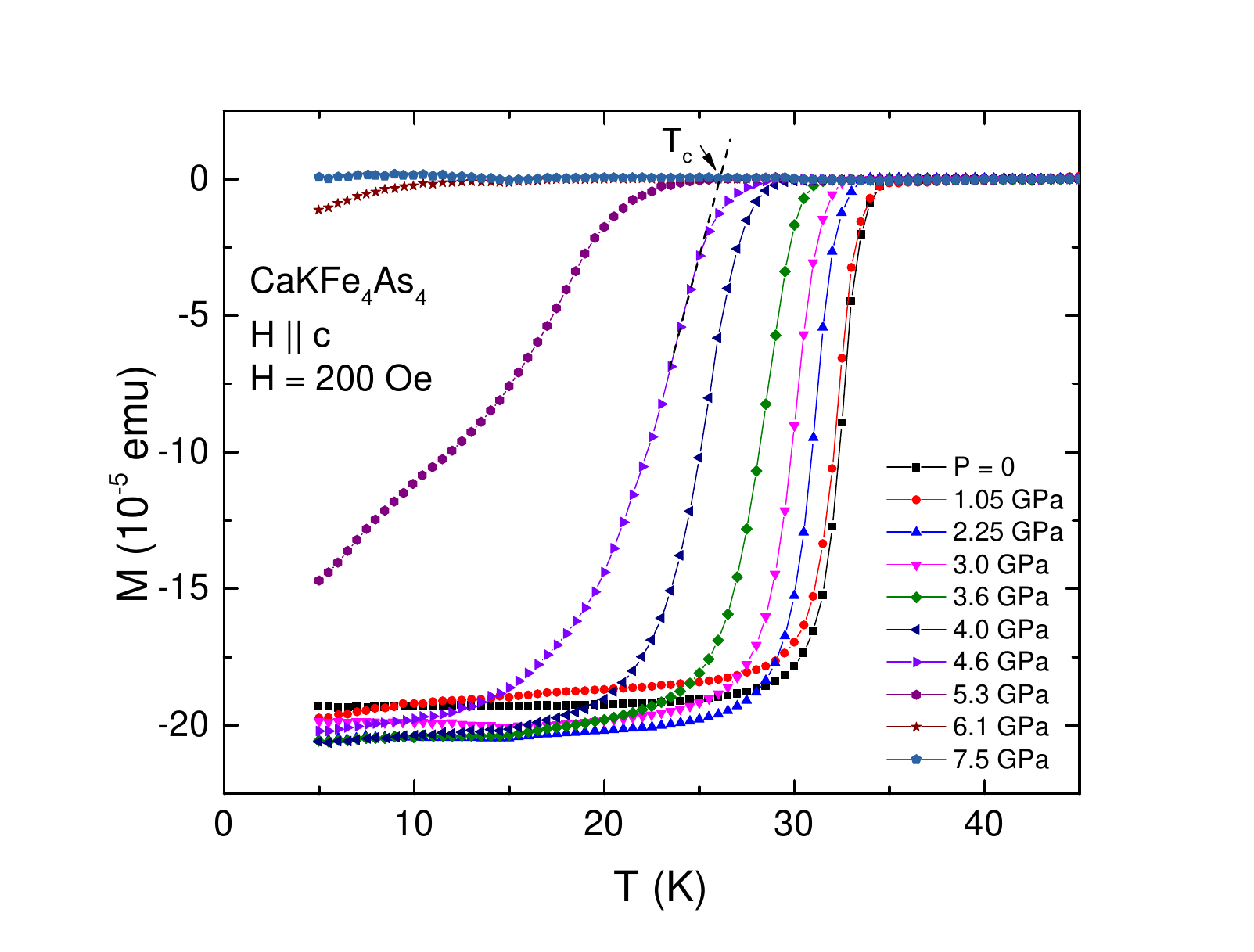}
\end{center}
\caption{(color online) Temperature dependent ZFC magnetization, $M(T)$ of  CaKFe$_4$As$_4$ crystal under pressure measured in 200~Oe magnetic field applied along the $c$-axis. The criterion for $T_c$ is shown using 4.6 GPa data. At $P > 4.6$~GPa superconductivity is apparently not bulk.} \label{F1} 
\end{figure}

To analyze in depth the impact of pressure on $T_c$, we compared our measured $T_c$ values with those previously reported in references \cite{kal17a} and \cite{wan23a}. The criterion for $T_c$ using $M_{trap}(T)$ is shown in detail in Appendix A. Figure \ref{F4} presents a summary of the data, demonstrating consistency between the values obtained from Fig. \ref{F1} and the literature. The data are consistent in terms of $T_c(P)$ and the critical pressure of the hcT, taking into account differences in samples, cells, and pressure media.

Within the Bean's model $M_{trap}(T)$ is proportional to $J_c(T)$. Typical low temperature self-field $J_c$ in thick CaKFe$_4$As$_4$ single crystals is around 2~MA/cm$^2$. \cite{hab19a} The main change under pressure in $M_{trap}(T)$ (ergo in $J_c(T)$) is the alteration in the functional dependence. To explain the shape of $J_c(T)$ at ambient pressure, as we mentioned in the Introduction section, a hypothesis involving two sources of pinning has been proposed: random disorder due to atomic chemical substitution \cite{fen18a} and temperature-activated pinning centers by stacking faults in superconducting CaFe$_2$As$_2$. \cite{ish19a}  Although bulk CaFe$_2$As$_2$ does not superconduct at ambient pressure \cite{nin08a}, it could superconduct in the form of intergrowths with ill-defined potentially strong inhomogeneous stresses and strains. \cite{tor08a,pro10a} According to this hypothesis the CaFe$_2$As$_2$ intergrowths contribute significantly to the pinning strength at higher temperatures when they are in a normal state but are excluded at lower temperatures when they are superconducting. This is suggested as an explanation for the broad hump feature in $J_c(T)$.  Following with this hypothesis further, it is not unreasonable to suppose that,  as in Ref. \cite{tor08a}, superconductivity in CaFe$_2$As$_2$ could be completely suppressed at some pressure, with both sources of pinning being active over the entire temperature range of CaKFe$_4$As$_4$ superconductivity. This would explain the "straightening" of $J_c(T)$ with pressure and would also help to understand the "flat" $M_{trap}(P) \propto J_c(P)$ at 5~K for more than 4~GPa (Fig. \ref{F2}, inset).  It is worth to point out that superconductivity in bulk CaFe$_2$As$_2$ under quasi-hydrostatic pressure is observed starting from $\sim 0.5$~ GPa and persists to about 1~GPa, \cite{tor08a} however superconductivity in the intergrown CaFe$_2$As$_2$ could be significantly different from the bulk. (Cf. superconductivity of intergrown Bi or Ni - Bi in single crystals of RNi$_{1-x}$Bi$_{2 \pm y}$, R = rare earth in Ref. \cite{lin13a}.) Whereas under pressure overall $J_c(T)$ is expected to decrease with the suppression of $T_c$ (e.g. see examples in Ref. \cite{san21a}), the low temperature pinning increases with growing contribution from normal state CaFe$_2$As$_2$ and low temperature $J_c$ is expected to increase . As a result a balance of these two opposite tendencies the base temperature $J_c$ is almost constant up to $\sim 4$~GPa.

The hypothesis of additional pinning centers at low temperatures in samples under pressure is confirmed by the observed slowing down of flux relaxation at 5~K, which correlates with the increase in low-temperature pinning discussed above. The vortex dynamics in iron-based superconductors is typically analyzed using the framework of collective pinning theory. \cite{pro08a} The reduction in flux creep is consistent with previous reports where additional pinning centers introduced by irradiation reduced the decay of persistent currents by enhancing the vortex pinning energy rather than altering the glassy exponent $\mu$ (typically around 1). \cite{hab20a} In our self-field results, the $S$ values start at approximately 0.016 and systematically decrease to around 0.006, similar to what is observed in samples with artificially introduced pinning centers or in single crystals with strong pinning by normal precipitates. \cite{yes96a,hab20b} In this context, CaKFe$_4$As$_4$ single crystals represent an exceptional system where, under external pressure, superconductivity is suppressed, but vortex pinning increases at low temperatures due to changes in microstructural defect properties at the nanoscale.

\section{Conclusions}

In summary, magnetic measurements of CaKFe$_4$As$_4$ crystals in DAC under pressure reveal a decrease of $T_c$ observed through low-field $M(T)$ and trapped flux magnetization measurements, as well as changes in  the temperature dependence of $J_c$ and slowing down of the 5~K flux creep rate. The latter two observations can be understood extending the hypothesis of two sources of pinning in CaKFe$_4$As$_4$ suggested in Ref. \cite{ish19a}. Above the hcT transition under pressure, when the sample is in non-bulk superconducting state, both, diamagnetism in $M(T)$ and trapped flux magnetization signal critically diminish and disappear. As final remark, CaKFe$_4$As$_4$ provide evidence of a strong competition between intrinsic superconducting properties and vortex pinning tuned by an external parameter such as pressure in a unique superconductor. As a final remark, CaKFe$_4$As$_4$ provides evidence of a strong competition between intrinsic superconducting properties and vortex pinning, both of which are tuned by an external parameter such as pressure, making it a unique superconductor to explore these kinds of behavior.

\ack

This work was supported by the U.S. Department of Energy, Office of Science, Basic Energy Sciences, Materials Sciences and Engineering Division. Ames National Laboratory is operated for the U.S. Department of Energy by Iowa State University under Contract No. DE- AC02-07CH11358. NH is a member of the Instituto de Nanociencia y Nanotecnolog\'ia INN (CNEA-CONICET) and was partially supported by PICT 2015–2171 (Argentina).

\clearpage

\clearpage

\clearpage

\clearpage

\begin{figure}
\begin{center}
\includegraphics[angle=0,width=150mm]{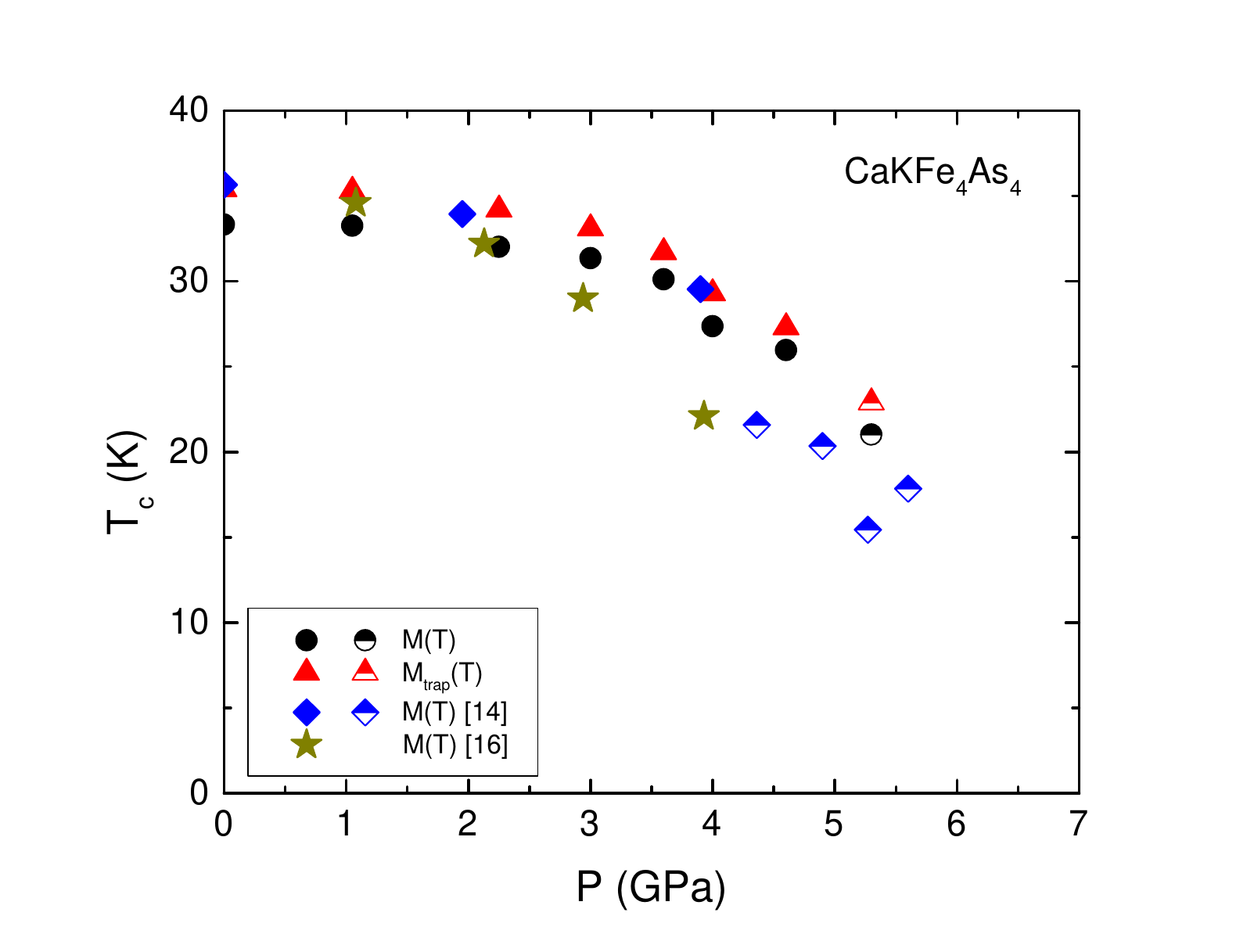}
\end{center}
\caption{(color online) Pressure dependence of the superconducting transition temperature from $M(T)$ and $M_{trap}(T)$ in this work and $M(T)$ in Refs. \cite{kal17a,wan23a}. Half-filled symbols correspond to the data showing potentially not bulk transition. } \label{F4} 
\end{figure}

\appendix

\section{$T_c$ criterion in trapped flux magnetization measurements}

\begin{figure}
\begin{center}
\includegraphics[angle=0,width=150mm]{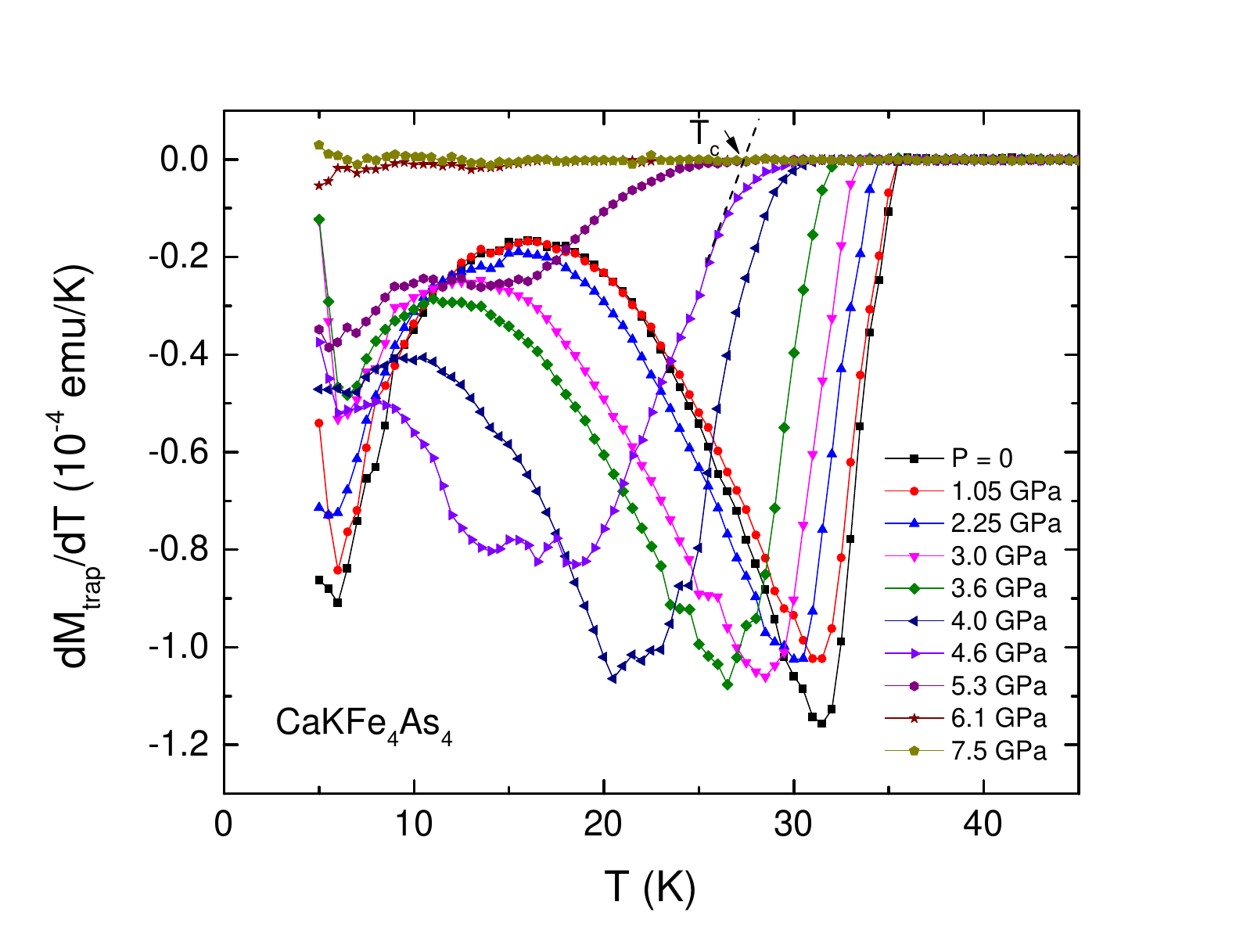}
\end{center}
\caption{(color online) Temperature derivatives of the trapped flux magnetization measurements at different pressures. For $P = 4.6$~GPa the $T_c$ criterion is shown.} \label{FA1} 
\end{figure}

Figure \ref{FA1} presents the temperature derivatives of the trapped flux magnetization measurements, $dM_{trap}/dT$, at different pressures. For $P = 4.6$~GPa the $T_c$ criterion is shown.

\clearpage

\end{document}